\documentclass[letter,keeplastbox]{jacow}
\usepackage[utf8]{inputenc}
\usepackage{amsmath}
\usepackage{siunitx}
\usepackage{float}

\ifboolexpr{bool{jacowbiblatex}}
 {
  \addbibresource{jacow-test.bib}
  \addbibresource{biblatex-examples.bib}
 }{}
\begin{document}

\title{Characterization of Octupole Elements for IOTA\thanks{Work partially supported by the US Department of Energy, Office of Science, High Energy Physics under Cooperative Agreement award number DE-SC0018362 and Michigan State University.}}

\author{J. N. Wieland\thanks{wielan22@msu.edu}, Michigan State University, East Lansing, MI 48823, USA \\
J. D. Jarvis, A. L. Romanov, A. Valishev, Fermilab, Batavia, IL 60510, USA}
	
\maketitle

\begin{abstract}
    The Integrable Optics Test Accelerator (IOTA) is a research storage ring constructed and operated at Fermilab to demonstrate the advantages of nonlinear integrable lattices. One of the nonlinear lattice configurations with one integral of motion is based on a string of short octupoles. The results of the individual magnet's characterizations, which were necessary to determine their multipole composition and magnetic centers, are presented. This information was used to select and align the best subset of octupoles for the IOTA run 4.
\end{abstract}

\section{Introduction}
\begin{figure}[h]
    \centering
    \includegraphics[width = 0.4 \textwidth]{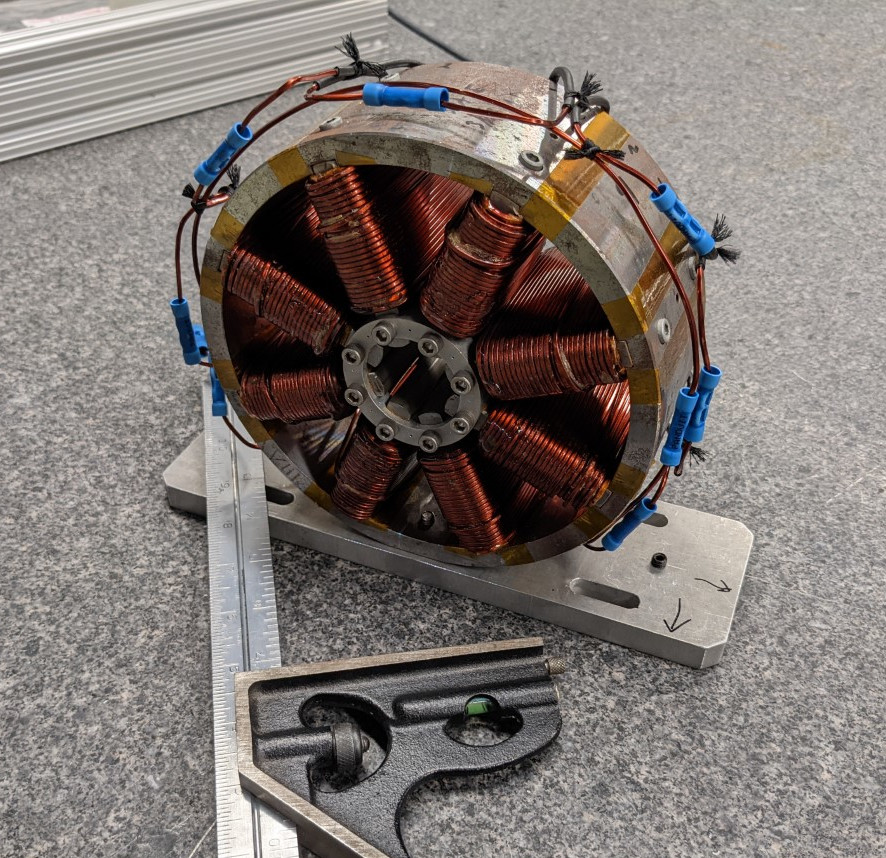}
    \caption{A single octupole from the string.}
    \label{fig:oct}
\end{figure}

As part of the Integrable Optics Test Accelerator (IOTA) a string of octupoles (Fig. \ref{fig:oct}) is installed in a configuration to maintain the Hamiltonian as a constant of motion. During IOTA run 2 unexpected deviations in the closed orbit while the octupoles were energized suggested misalignment in the magnets or deviations in construction generating large low-order (quadrupole and dipole) transverse multipole components. The nominal values for the octupoles are in Table~\ref{tab:octParams} \cite{antipov2016design}.

There are a number of conventions for presenting the multipole components, and the following format will be used in this paper, Eqs. (\ref{eq:harmDef}) \& (\ref{eq:compDef}).

\begin{equation}
    B_y + iB_x = \sum_{n=1}^\infty C_n \left(\frac{x+iy}{R_{ref}}\right)^{n-1}
    \label{eq:harmDef}
\end{equation}

\begin{equation}
    C_n = B_n + iA_n
    \label{eq:compDef}
\end{equation}
Where $B_n$ and $A_n$ are the normal and skew terms respectively, $R_{ref}$ is the reference radius for the measurements, and the multipole index "$n$" follows the European convention, i.e. $n$ = 1 corresponds to the dipole term. The longitudinal component of the field was not considered in the characterization. The magnets were removed and characterized using a hall probe to determine potential outliers and align a set of nine magnets for installation in a new configuration before IOTA run 4. The figure of merit for selecting the magnets was the magnitude of low order multipoles.

\begin{table}[h]
    \centering
    \caption{Nominal Octupole Parameters}
    \begin{tabular}{@{}lc@{}}
        \toprule
        \textbf{Octupole Parameter} & \textbf{Design Value}\\
        \midrule
        Length & \SI{70}{mm} \\
        \midrule
        Aperture & \SI{28}{mm} \\
        \midrule
        Coil Turns per Pole  &  88 \\
        \midrule
        Maximum Excitation Current & \SI{2}{A} \\
        \midrule
        Maximum Octupole Gradient & \SI{1.4}{kG/cm^3} \\
        \midrule
        Effective Field Length & \SI{75}{mm} \\
        \bottomrule
    \end{tabular}
    \label{tab:octParams}
\end{table}

\section{Test Stand Measurements}
\subsection{Methods}

\begin{figure}[h]
    \centering
    \includegraphics[width = 0.35 \textwidth]{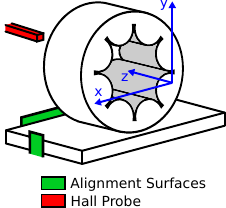}
    \caption{Cartoon of hall probe and support stand with octupole.}
    \label{fig:testCartoon}
\end{figure}

The multipole components of the magnets were determined using a hall probe mounted on a three-axis test stand based on a procedure described in reference \cite{campmany2014determination}. The test stand was composed of three, perpendicular rails actuated by linear stepper motors with a hall probe mounted along the nominal z-axis. The magnets were mounted to a support stand with alignment features for all degrees of freedom next to the test stand, see Fig. \ref{fig:testCartoon}. Before any measurements were taken, the test stand was calibrated to the support stand stand using a precise flat and dial indicator to ensure that the axes of motion were perpendicular to each other. 

All measurements were taken at an energizing current of \SI{2}{A}, the maximum for these octupoles. The test stand measured the magnetic field at a preprogrammed set of points. In practice, this was an equidistant set of points on a circle in a number of planes along the magnet's axis. The field from the x and y hall sensors was combined into azimuthal data based on the relative angle of the points on the circle. A Fourier decomposition was then performed on the magnetic field data to find the multipole components. A coarse scan (smaller radius, fewer points) was performed first and the relevant offset was calculated using Eq. (\ref{eq:octCenter}) assuming that the sextupole component was all due to feed-down. The probe would then be centered in the magnet based on this offset and proceed onto a second, higher-fidelity scan. The high fidelity scan was 32 points at a reference radius of \SI{8}{mm}, the largest radius which did not risk hitting any pole tips. In total, six circular scans were performed in the magnet at three different longitudinal positions, at each end of the pole tips and at the center of the magnet so the integrated field could be calculated. The measurements were taken moving forward and backwards through the magnet and averaged at each position to account for any potential backlash in the test stand. 

\begin{equation}
    x_o + i y_o  = \left(\frac{1}{n-1}\right)\left(\frac{C'_{n-1}}{C'_n}\right)R_{ref}
    \label{eq:octCenter}
\end{equation}

Once the magnets multipole compositions had all been determined the best magnets could be selected. As the sextupole components had been deliberately minimized, the quadrupole and dipole components were used for selecting the best subset. Any outliers were excluded and from the remaining ten magnets were selected (nine for installation and one spare). These magnets were then remeasured on the stand for alignment. The same basic procedure was followed, but the probe left in the same position for each magnet. The relative magnetic centers could then be calculated by Eq. (\ref{eq:octCenter}) and the magnets shimmed against matching alignment surfaces on the installation mount.

\subsection{Results}
\begin{figure}[h]
    \centering
    \includegraphics[width = 0.45 \textwidth]{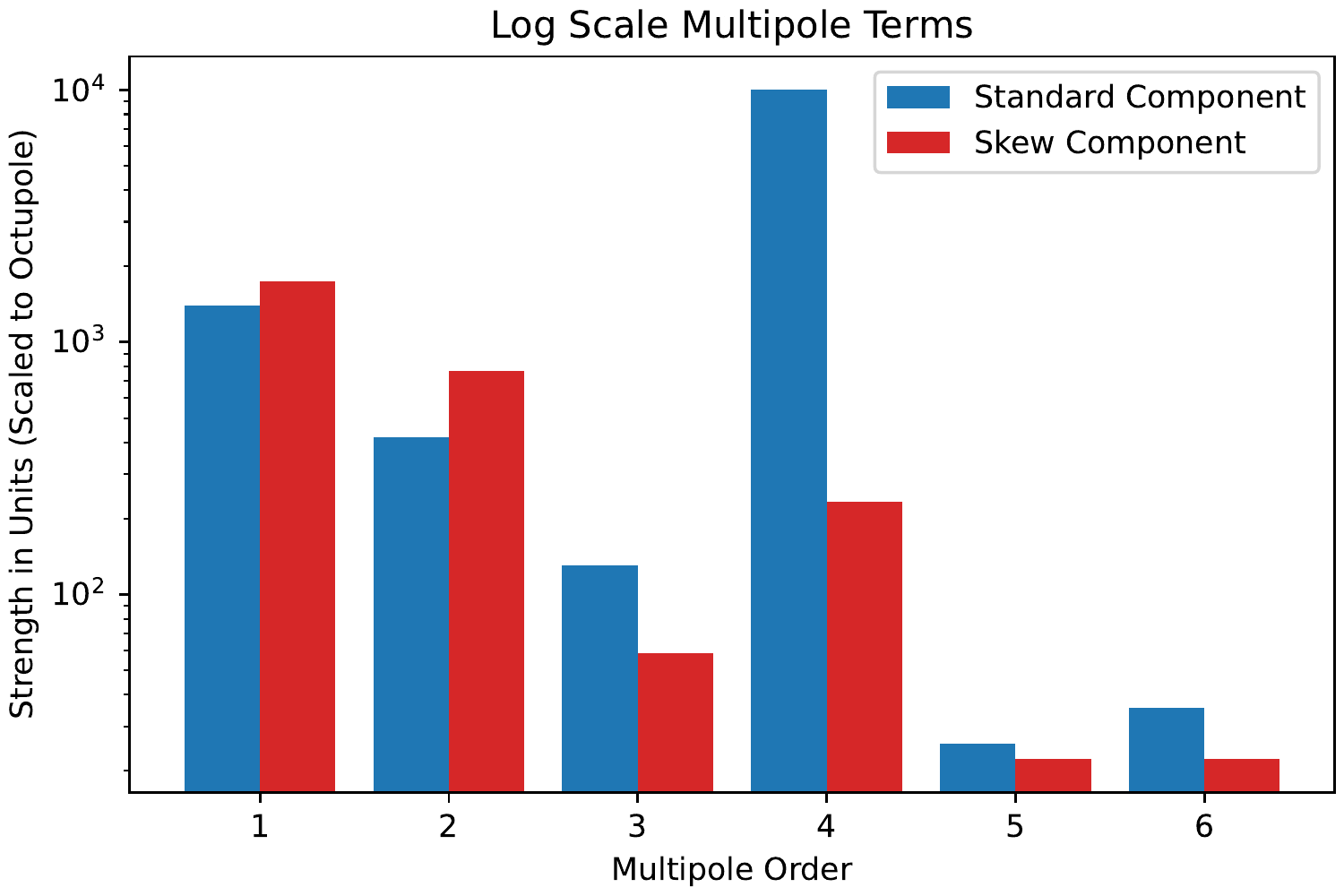}
    \caption{Initial multipole decomposition.}
    \label{fig:azDecomp}
\end{figure}

Initially, all magnet decompositions demonstrated abnormally large low-order multipole components, especially the dipole term (see Fig. \ref{fig:azDecomp}).

This did not match empirical measurements, the abnormal dipole term was similar for all longitudinal positions in the magnet, but fields of this magnitude were not observed at the center where higher order multipoles are negligible. The source of this error was found to be the use of the calculated azimuthal fields. The hall probe consist of three individual sensors in the probe tip and have a significant offset with respect to one another. The initial calculation assumed these measurements were taken at the same point in the probe. To remedy this, the individual measurements of the disparate hall probes were decomposed separately, so each pass on the test stand effectively took X and Y measurements. The longitudinal component was not used. These measurements were not aligned to the magnetic center as the centering movement of the probe was based on the azimuthal calculation. In the interest of time, the measurements were centered in software using Eq. (\ref{eq:recenter}) which applies the feed-down of all higher order multipoles to find the components in a new set of coordinates \cite{jain1997basic}. As the sampling rate of the circular scan was 32 points the discrete Fourier transform yielded multipole components up to n=16, but the n=8 multipole is the maximum which demonstrates good sensitivity. A comparison of the centering calculation was done with both up to n=16 and n=8 and no significant deviations were observed.

\begin{equation}
    C_n = \sum_{k=n}^{\infty} C'_k \left(\frac{(k-1)!}{(n-1)!(k-n)!}\right)\left(\frac{x_o +i y_o}{R_{ref}}\right)^{k-n}
    \label{eq:recenter}
\end{equation}

\begin{figure}[h]
    \centering
    \includegraphics[width = 0.45 \textwidth]{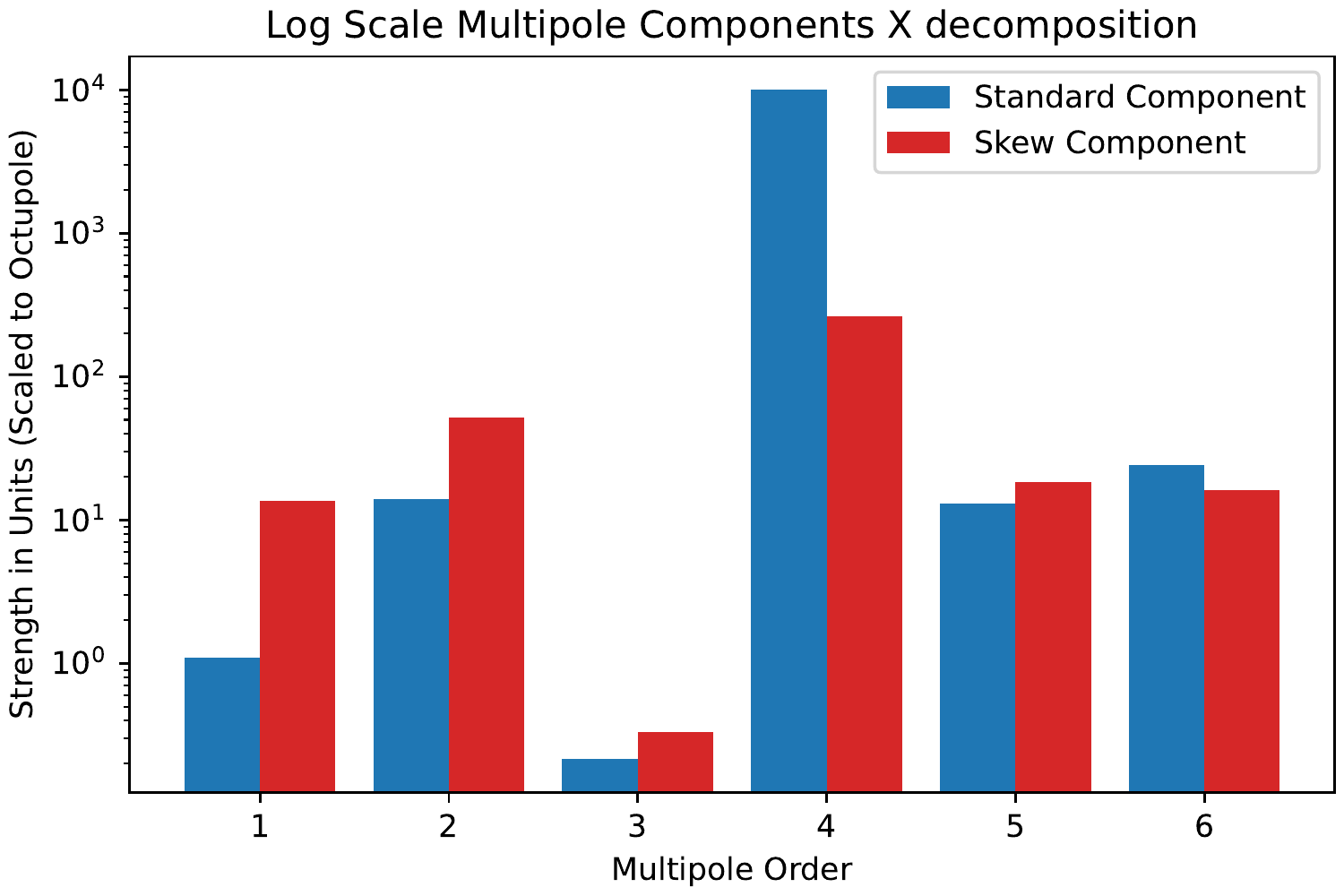}
    \caption{Multipole components from X sensor.}
    \label{fig:xDecomp}
\end{figure}

\begin{figure}[h]
    \centering
    \includegraphics[width = 0.45 \textwidth]{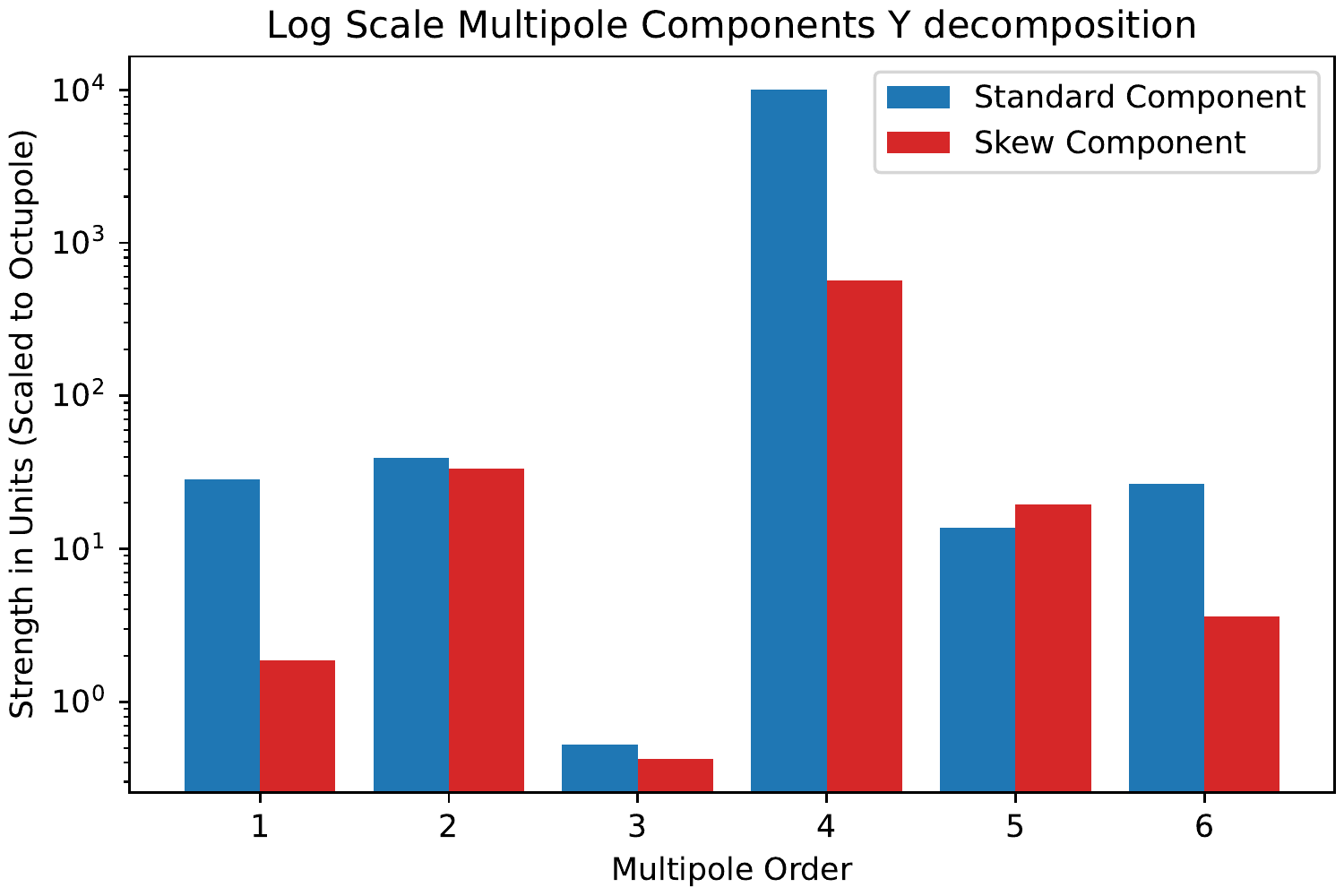}
    \caption{Multipole components from Y sensor.}
    \label{fig:yDecomp}
\end{figure}

The new decomposition yielded much cleaner results (Figs. \ref{fig:xDecomp} and \ref{fig:yDecomp}), and was used for the centering measurements. 

\subsection{Alignment}
\begin{figure}[h]
    \centering
    \includegraphics[width = 0.48 \textwidth]{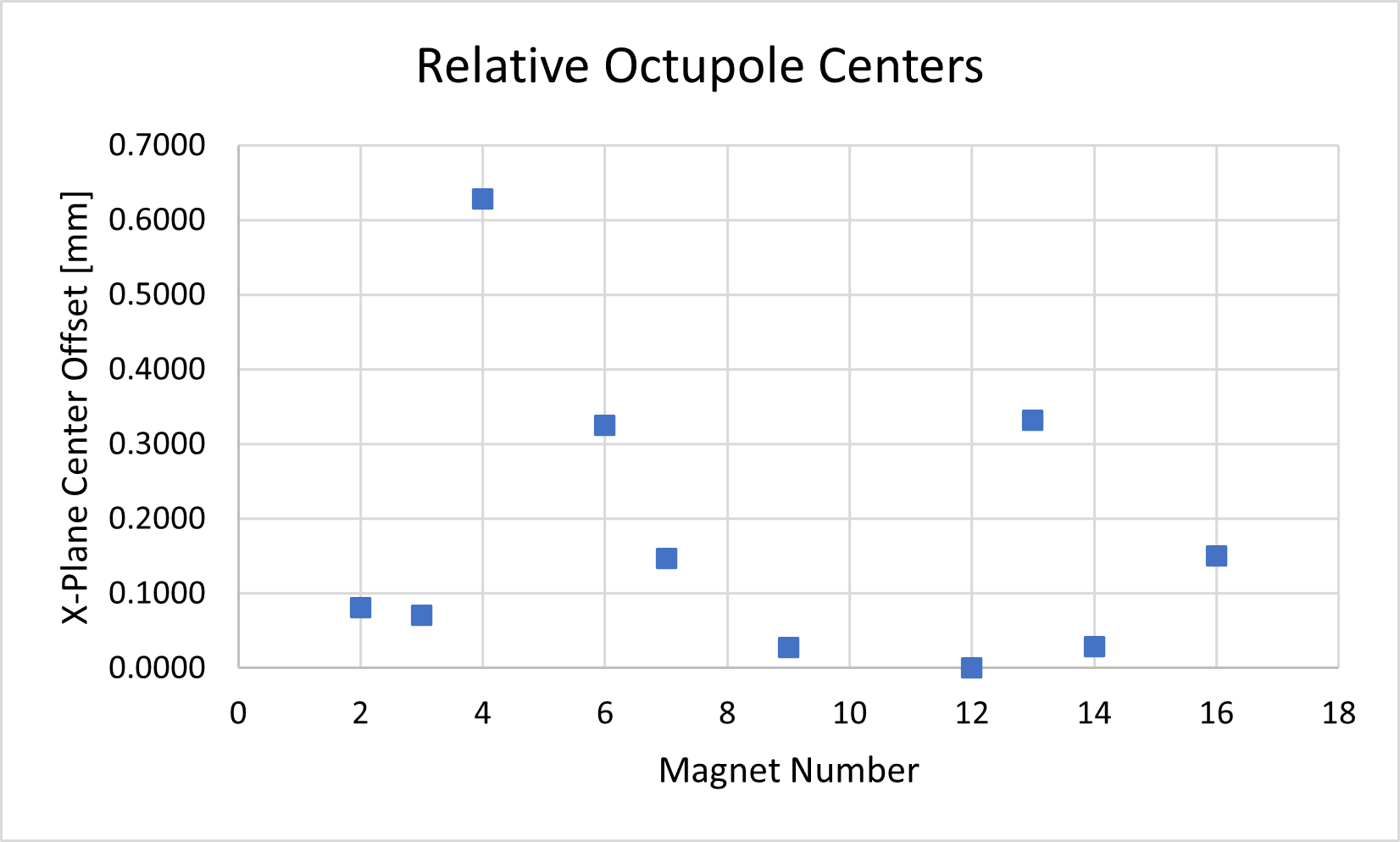}
    \caption{Relative offset of octupole magnets.}
    \label{fig:xOctCenters}
\end{figure}

Once a subset of octupoles had been selected the alignment measurement was performed. During the course of these measurements the repeatability of the test stand positioning at the center was found to be on the order of \SI{5}{\micro m}. This was well within the alignment threshold of \SI{400}{\micro m}. To determine the size of the shims, the relative offset compared to the magnet with the center furthest from the alignment feature was found. Figure \ref{fig:xOctCenters} shows the offsets in the x direction for the selected magnets, here Magnet 12 is the reference.

\begin{figure}
    \centering
    \includegraphics{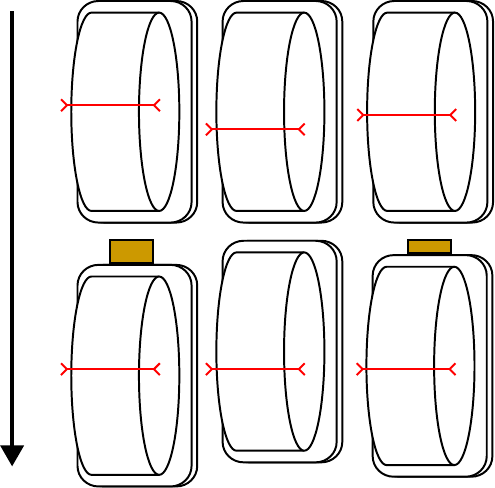}
    \caption{Cartoon of relative alignment procedure.}
    \label{fig:octAlign}
\end{figure}

Shims matching these offsets could then be inserted to align the relative centers of the octupoles (see Fig. \ref{fig:octAlign}).

\section{Summary and Future Work}
The octupoles for the IOTA quasi integrable lattice element were characterized using a hall probe mounted to a test stand. An error related to the mismatched position of the hall sensors in the probe was identified and remedied using an alternative decomposition of the field. Once the satisfactory magnets had been selected, their relative centers were measured and aligned for installation in IOTA prior to run 4. In the course of run 4, the magnet alignment will be confirmed using beam based measurements.


\end{document}